\documentclass[prdnotes,showpacs,12pt]{revtex4}
\pdfoutput=1
\usepackage[latin1]{inputenc}
\usepackage{graphicx}
\usepackage{amsmath}
\usepackage{amssymb}
\usepackage{braket}
\DeclareMathAlphabet{\mathpzc}{OT1}{pzc}{m}{it}

\numberwithin{equation}{section}

\newcommand{\RomanNumeralCaps}

\makeatletter
\gdef\@fpheader{}
\makeatother

\begin{document}

\title{On the momentum of gluons in Lattice Gauge Theory (LGT)}
\author{\bf Herbert Neuberger}
\email{herbert.neuberger@gmail.com}
\affiliation{ Department of Physics and Astronomy, Rutgers University\\Busch Campus, Piscataway, NJ088540}


\vspace{2.5mm}

\begin{abstract}
{Attempts to improve LGT simulation algorithms by Fourier space preconditioning have been handicapped by the gauge dependence of momenta, familiar from perturbation theory. The continuum theory has a gauge invariant energy-momentum density, indicating this to be a fake obstacle. While perturbation-theory  momenta are evidently wrong for the task, momentum-transfer carried by gluons is physical. Simulations using these momenta may become practical in the near future thanks to recent progress in generative models, stochastic and/or deterministic.}
\end{abstract}

\maketitle

\section{Introduction}

Particle Physics is described by a field theory, defined by a Lagrangian whose properties evolve with scale  by integrating out fields carrying higher momenta and introducing new lower momentum field variables in order to obtain a practical effective Lagrangian which adequately describes the Physics at lower scales.  

It is straightforward to carry out this momentum 
renormalization group (RG) step in perturbation theory. Beyond perturbation theory, in LGT, local averaging could separate the lower energy variables from the integrated out ones, but implementation is awkward. 

In practice, LGT projects typically forgo the RG altogether, applying local stochastic algorithms to all fields. Results of LGT must then be connected to perturbative formulas. Generic numerical work exploits Lagrangian locality in space, but the latter ceases to be evident in momentum space. Relatively recent progress in generative modeling, stochastic or deterministic\cite{van}, seems to be well positioned to overcome this problem now and motivates this short paper. 

More theoretical pursuits by Sasha Migdal\cite{mig} have already a long time ago led him to express lattice loop equations in momentum space. They involve objects that often are displayed by identical sketches in both momentum and real spaces. Migdal's form of his momentum variant raised hope for a better road to a solution.

\section{Gauge invariant conserved momentum in LGT.}

Lattice perturbation theory is done in momentum space, necessarily with gauge fixing, mirroring continuum. This momentum is carried by the lattice vector potential, ${\hat  A}_{x,\mu}$, the Fourier Transform (FT) of the antihermitian exponent of the $U_{x,\mu} \in SU(N)$ associated with the link $x \to x+\mu$. 

The prospect of easy efficient preconditioning 
\cite{george} of simulation algorithms in momentum space for matter fields (exploiting the form of derivative terms in the action) motivated extension to gauge fields, where gauge fixing is neither necessary nor advisable, but 
was nevertheless adopted, becoming a major obstacle beyond perturbation theory, both in principle\cite{brs2} and in practice. 

The next section eliminates this obstacle first, trying to avoid a doomed-to-failure gauge-fixed program. The continuum limit must have a well defined, gauge invariant local energy momentum tensor: hence, in principle gauge dependence must be avoidable. In addition, Migdal's momentum-space loop equations had already shown that gauge fixing isn't mandatory. 

\subsection{Parallel transport.}

Starting from continuum, it will be shown below that Wilson's LGT can be formulated in terms of matrices containing only gauge invariant momenta.

Consider a smooth and flat four dimensional torus carrying 
an $SU(N)$ Yang Mills field with standard continuum action in the formal path integral. In LGT the path integral is replaced by a sequence of well defined integrals over fields restricted to a finite four dimensional lattice embedded in the continuum torus and coherently oriented with it. Both have a four dimensional hyper-cubic symmetry. The side of the continuum torus has length $l$ and the corresponding number of lattice sites is $L$. Continuum is obtained in a limit where the lattice spacing $a\to 0$ ($a = l/L$) in a way correlated with an adjustment of the coupling so that $l$ becomes a finite physical length in the limit. There exists no rigorous proof of this limit's existence, but checks over the last $\sim 50$ years have been consistently more convincing as time passed and by now doubts have disappeared from the mainstream.

In continuum parallel transport occurs along an open curve
${\mathcal C}$ connecting $x \to y$. It is defined by integration over infinitesimal steps along it. What gets transported is a vector in ${\mathcal V}_x \cong C^N$. Both
in continuum and on the lattice ${\mathcal V}_x$ and ${\mathcal V}_y$ are distinct whenever $y\ne x$: there is no relationships between basis choices at different locations. ${\mathcal C}$ connects
$x_0\to x_1$ and parallel transport is a unitary map 
${\mathcal V}_{x_0}  \to {\mathcal V}_{x_1}$. Parallel transport from $x_1 \to x_0$ along the reversed ${\mathcal C}$ is by definition the inverse map. Parallel transport provides an intrinsic structure on the totality of spacial manifold and vector spaces using a provided connection $A_\mu$ associated to an operator $D_\mu \equiv (\partial_\mu + iA_\mu)$ with $A_\mu \in su(N)$, the Lie algebra  of $SU(N)$. The desired map ${\mathcal V}_{x_0} \to {\mathcal V}_{x_1}$ consists of consecutive infinitesimal steps along ${\mathcal C}$, defined by the first term in $D_\mu$, ($\partial_\mu$) and unitaries close to one defined by the second, ($A_\mu$) acting on bases in ${\mathcal V}_z$ as $z$ moves along  ${\mathcal C}$ from $x_0 \to x_1$. The combined unitary is the map and the collection of all these maps synchronize the entire construct.  

The infinite number of $A_\mu (x)$'s are the formal dynamical integration variables of the continuum path integral. The lattice replaces this ill defined continuum integration by a proper integral, restricting the sites $x$ to a finite torus and the set of paths ${\mathcal V}_x \to {\mathcal V}_y$ to ones made out of head to toe fused single links. The connection is replaced by the bounded $e^{iaD_\mu}$ which are the lattice elementary parallel transporters $T_\mu$, consisting of finite unitary matrices destined to be integrated over. Like the 
$D_\mu$ operators, the $T_\mu$ matrices act on all of $\bigoplus_x {\mathcal V}_x$. For each $x$ every  $v_x \in  {\mathcal V}_x$  is mapped to $v_{x+\mu}
\equiv U_{x+\mu}v_x \in {\mathcal V}_{x+\mu}$, with $U_{x+\mu} \in SU(N)$ the familiar independent link matrices on $x\to x+\mu$, integrated under Haar measure augmented by the usual action weighting of standard LTG. The preceding discussion identifies the operators $T_\mu$ as finite  lattice matrix versions of the continuum operators $e^{iaD_\mu}$.

FT acts on  $T_{\mu}$ by conjugation replacing site-index pairs with momentum-index pairs. If we write $T_\mu$ in the exponential form for a unitary matrix, the FT acts on its exponent again by conjugation, leaving it dependent only on momenta-index differences, i.e. momentum transfers, a specific property of FT. They represent physical forces between interacting gauge singlet
particles, obviously gauge invariant, conserved, physical and fit to separate hard from soft gauge fields in a RG scheme. The momentum space matrices entries
with identical momentum index differences are equal to each
other. Hence, there is a unique gauge invariant momentum transfer associated to any $T_\mu$, a structural restriction, clearly necessary physically. An unavoidable penalty is incurred: whereas the standard site-based Wilson formulation has a simple factorized integration measure, here the measure is more complicated. The modern results I alluded to earlier as my motivation appear to be notably good at implementing awkward integration measures exploiting specifically designed deterministic or stochastic connecting deformation flows between the simple and complicated probability distributions. Setting this aside, there is no hindrance to defining Wilson's LGT by
$T_\mu$'s carrying physical momenta.

\subsection{Momentum space and gauge invariance}

Continuum $A_\mu$  transforms inhomogeneously under gauge transformations, invalidating the naive RG prescription of integrating out high momentum carrying $A_\mu$ fields. Doing this will lead to undesirable gauge breaking in the lower momentum effective Lagrangians. Gauge theories have a gauge invariant energy-momentum tensor and the conserved quantity is momentum \underline{transfer}. An elastic scattering of two physical particles has a measurable momentum transfer which is the balance of the transfer momenta carried by gluon exchanges. $A_\mu$-FT-momentum evidently is not the same as $T_\mu$-FT-momentum: in principle (and hopefully in practice) it is adequate for FT acceleration. 

\section{Final comments.}
      I first became acquainted with the usefulness of the $T_\mu$ objects in \cite{bnds} which obtained an improvement of a bound relevant to the overlap fermions' construction which, to my knowledge, has not been improved since then. Earlier reference to such objects can be found in \cite{vanbaal}. Similar objects appeared earlier in \cite{orland},  in the context of large $N$ reduction. I have used them recently in \cite{planar} to provide a complete proof for the quenching prescription  of  \cite{BHN} directly in lattice perturbation theory\cite{planar}. The $T_\mu$'s in the proof implied a doubling  of the index types inhabiting each of the two members of `t Hooft's  famous double lines for planar diagrams. More technical details  appeared in \cite{UY}. In the present  paper I emphasized the naturalness of the $T_\mu's$ from the standard mathematical viewpoint of gauge theory.
 
 The $T_\mu$'s would be useful in other applications, including perturbation theory, but an implementation employing generative modeling is more timely now.  

 {\bf No-brainer conclusion: \framebox{For gauge invariant $p_\mu$ use LTG $D_\mu$.}}
   
 \end{document}